\documentclass[11pt]{article}
\PassOptionsToPackage{numbers}{natbib}
\usepackage{acl}
\usepackage{times}
\usepackage{latexsym}
\usepackage[T1]{fontenc}
\usepackage[utf8]{inputenc}
\usepackage{microtype}
\usepackage{inconsolata}
\usepackage{graphicx}
\usepackage{array}
\usepackage{amsmath}
\usepackage{amssymb}
\usepackage{algorithm}
\usepackage{algorithmic}
\newif\ifshowtodos
\showtodosfalse 

\ifshowtodos
  \usepackage[textsize=tiny,backgroundcolor=pink, ]{todonotes}
\else
  \usepackage[disable,textsize=tiny,backgroundcolor=pink]{todonotes}
\fi
\ifshowtodos
  \newcommand{\reply}[1]{\par\vspace{1ex}\noindent\rule{\linewidth}{0.3pt}\par\vspace{1ex}\noindent{\color{blue!60!black}\textbf{Essam:} #1}}
\else
  \newcommand{\reply}[1]{}
\fi
\usepackage{caption}
\usepackage{placeins}
\usepackage{float}
\usepackage{needspace}
\usepackage[most]{tcolorbox}
\usepackage{titletoc}
\usepackage{enumitem}
\usepackage{xspace}
\usepackage{bm}
\usepackage{fontawesome5}
\usepackage{xurl}

\title{KGVoyager: Knowledge Graph Agnostic Question Answering via Agentic Navigation}

\author{
  Essam Wisam \and Chengkai Li \\
  University of Texas at Arlington \\
  \texttt{\{abdelghanye, cli\}@uta.edu}
}

\begin{document}
\newcommand{\entity}[1]{\texttt{\small #1}}
\newcommand{\rel}[1]{\textsl{\small #1}}
\newcommand{\scriptsizerel}[1]{\textsl{\scriptsize #1}}
\newcommand{\triple}[3]{(\entity{#1}, \rel{#2}, \entity{#3})}
\newcommand{\entityname}[1]{\textsf{\footnotesize #1}}
\newcommand{\relname}[1]{\textsf{\emph{\small #1}}}
\newcommand{\entityt}[1]{\texttt{\scriptsize #1}}
\newcommand{\tinytriple}[3]{\entityt{#1}, \rel{#2}, \entityt{#3}}

\newcommand{\bsystem}[1]{{\small\ensuremath{\bm{\mathsf{#1}}}}\xspace}
\newcommand{\bkgv}{\bsystem{KG Voyager}}

\newcommand{\sbsystem}[1]{{\scriptsize\ensuremath{\bm{\mathsf{#1}}}}\xspace}
\newcommand{\sbkgv}{\sbsystem{KG Voyager}}

\newcommand{\tsystem}[1]{{$\mathsf{#1}$}\xspace}
\newcommand{\tkgv}{\tsystem{KG Voyager}}

\newcommand{\system}[1]{{\small \ensuremath {\mathsf{#1}}}\xspace}
\newcommand{\kgv}{\system{KG Voyager}}
\maketitle

\begin{abstract}
Knowledge Graph Question Answering (KGQA) over RDF graphs remains challenging in domain-specific settings, where formal ontologies and curated text–SPARQL pairs are often unavailable. We present \kgv, a KG-agnostic agentic architecture that generates SPARQL queries from natural language questions by dynamically discovering graph structure and semantics, requiring only a query endpoint of the underlying graph. Using a think–act–observe loop with search, exploration, and execution tools, \kgv\ maps terms to graph Internationalized Resource Identifiers (IRIs), uncovers structure, and refines queries through execution feedback—all without pre-existing ontologies or examples. Unlike the prior state of the art, \kgv\ requires only a lightweight class index which renders it applicable for far more real-world endpoints. Across four benchmarks, \kgv improves F1 by ${\sim}$4 points on average compared to the state of the art while cutting cost by ${\sim}$38\% and runtime by ${\sim}$24\%. 
\end{abstract}

\section{Introduction}
\label{sec:introduction}

RDF knowledge graphs (KGs) provide a robust framework for representing complex, interconnected data across diverse domains \cite{hogan2021knowledge,turki2024decade}. However, utilizing them effectively requires understanding graph-specific data models and technical query languages such as SPARQL \cite{jian2025interacsparql,ferre2013squall,li2024knowledge, abusalih2021domain}. 
Natural language interfaces that map user questions to SPARQL queries lower this barrier and have driven significant research interest in knowledge graph question answering (KGQA) \cite{dabramo-etal-2025-investigating, gashkov2025sparql, liu2024spinach,zahera2024generating,sgpt2022generative,kacupaj2023sparqlgen}.\todo[color=cyan]{how do we discuss the connection between QA and SPARQL queries. Ideally our study might want to use 3 different types of ``queries'': 1. bag of keywords; 2. QA; 3. textual specification of structured queries such as SPARQL.}

LLM-based KGQA systhems have achieved state-of-the-art benchmark performance~\cite{dabramo-etal-2025-investigating} over earlier machine-learning~\cite{berant2013semantic}, deep-learning~\cite{neural2022machine, sgpt2022generative, banerjee2022modern}, template, and rule-based approaches~\cite{ferre2013squall}. These solutions fundamentally leverage ontologies and example text--query pairs for grounding and guiding query generation. 
Ontologies encode classes, attributes, relationships, and constraints that help models produce semantically valid queries \cite{emonet2024llm,mishra2022natural,parot2020}. In addition to ontologies, example text--SPARQL pairs improve performance of such natural language interfaces by serving as in-context examples or facilitating fine-tuning \cite{wikisp2023,zahera2024generating,sun2022think}. This dual reliance on ontologies and text--query pairs parallels SQL-based DBQA, where systems depend on the database schema and text--SQL pairs to ground query construction \cite{llm_text_to_sql_survey_2024}. 

However, the ontologies and example text–SPARQL pairs on which these systems rely are not always available for domain-specific knowledge graphs. We identify three distinct challenges. (1) \emph{Ontology deficiency}: Unlike relational databases, which require explicit schemas, RDF permits a flexible, schema-optional representation. Many knowledge graphs lack a formal ontological structure. They may define entities, classes, and properties without specifying pivotal semantic relationships such as class hierarchies, property characteristics, and other semantic attributes. (2) \emph{Ontology opacity}: Even when knowledge graphs do incorporate ontological structure, they predominantly lack documentation that exposes or explains such structure and related constraints~\cite{graphbuild2023ontology}.\todo[color=cyan]{This reference doesn't appear to explain the opacity; We also don't have reference for deficiency; later, the FRINK website reference doesn't establish evidence for scarcity either. In general, we lack evidence in this paragraph.

\reply{
    From the website, multiple quotes support deficieny including:
    "It is possible to have a graph model without an ontology. In fact, many graph databases and graph-based applications..."

    As for opacity, I don't have a reference but it's hard for someone to dispute the claim by implying that documentation often exists. Manually checking FRINK KGs supports this claim.

    Lastly, for example scarsity the other reference mentions it clearly. FRINK is only a reference for "domain specific KGs". Can be removed if you find it appropriate.
}
} (3) \emph{Example scarcity}: Creating text–SPARQL example pairs for new domain-specific knowledge graphs is both costly and challenging, requiring human experts with deep domain knowledge, advanced SPARQL proficiency, and a thorough understanding of the underlying ontology. This is corroborated by the pronounced scarcity of query examples for domain-specific knowledge graphs~\cite{rangel2024sparql,frink2026}.

These challenges hinder the development of effective KGQA systems for \emph{domain-specific} settings. Most recent natural language to SPARQL systems~\cite{liu2024spinach,zahera2024generating,sgpt2022generative,kacupaj2023sparqlgen,banerjee2022modern,diallo2025frase,sun2022think,wikisp2023} avoid these constraints by focusing on \emph{open-domain} knowledge graphs such as Wikidata and DBpedia, which benefit from inherent knowledge in LLMs~\cite{gashkov2025sparql,wikisp2023}, curated text--SPARQL datasets~\cite{dubey2019lcquad, usbeck2019qald,usbeck2021qald}, and community ontology tooling \cite{liu2024spinach,zahera2024generating}. These assets are rarely available for arbitrary domain-specific knowledge graphs. 

As ontologies may be unavailable or opaque and example pairs scarce, we treat a SPARQL query endpoint as the sole universally assumable resource. Our system aims to enable natural language interfaces for arbitrary knowledge graphs \emph{given only query endpoint access}. To achieve this goal, our system must simulate the exploratory learning process a human expert would undertake when encountering an unfamiliar knowledge graph. To this end, we present \kgv, an agentic architecture that uses a suite of KG-agnostic tools to dynamically discover latent graph structure, semantics, and instance data. \kgv then uses this discovered information to generate SPARQL queries that answer user questions. We design a set of searching, sampling, and navigation tools that enable systematic exploration of arbitrary knowledge graphs. Using these tools within a think-act-observe loop \cite{yao2023react}, the agent identifies relevant classes and properties, maps local graph structure, and validates query drafts through execution feedback until it derives an accurate SPARQL query.

\kgv generalizes across endpoints by relying on two assumptions. First, the knowledge graph follows the W3C RDFS standard for defining types~\cite{w3c_rdf_schema}, so that class discovery and entity search can be expressed as agentic tools that run unchanged on any such endpoint. Second, a lightweight class index, containing the list of entity types in the graph, can be retrieved from the endpoint with a single SPARQL query; individual entities are then located on demand within the relevant classes. Both assumptions hold broadly in practice: in a scan of knowledge graph endpoints from the Open Knowledge Network (OKN) registry~\cite{frink2026}, $\sim$96\% of the scanned knowledge graphs follow the RDFS typing standard (Appendix~\ref{sec:appendix_rdf_typing}), and the class index can be built for $\sim$68\% of the tested endpoints under the Apache Jena SPARQL engine and $\sim$96\% under the QLever SPARQL engine (Section~\ref{sec:feasibility}). 

To the best of our knowledge, the only prior work sharing a similar endpoint-only, KG-agnostic objective is GRASP~\cite{walter2025grasp}, which also targets zero-shot SPARQL generation over arbitrary knowledge graphs. Unlike \kgv, which requires only an index of the graph's classes, GRASP assumes that an index covering all entities in the graph can be retrieved from the endpoint. This assumption holds for only $\sim$21\% of the tested OKN endpoints under Jena and none under QLever. Other fundamental differences are detailed in Section~\ref{sec:related}. Even when GRASP is given its full entity index, \kgv outperforms it on average across four domain-specific benchmarks, with $\sim$4 points higher F1, $\sim$38\% lower cost, and $\sim$24\% shorter runtime (Section~\ref{sec:evaluation}).

In summary, our key contributions are:\todo[color=cyan]{Some reviewers will pick on not seeing methodology novelty. So it will benefit ourselves if we could manage to point out novel aspects of the methodology. As of now it is easy for a reviewer to label this as "engineering" work.}
\begin{itemize}[noitemsep,wide,topsep=0pt]
    \item  \textbf{KGVoyager}, an agentic architecture that leverages a suite of KG-agnostic tools to generate SPARQL queries that answer user questions without relying on pre-existing ontologies or curated example pairs. 
    \item \textbf{Evaluation} of \kgv against the current state-of-the-art across diverse domain-specific knowledge graphs, demonstrating superior performance in correctness, efficiency, and applicability.
    \item \textbf{Open-source release} of \kgv, enabling practitioners to deploy the system against any SPARQL endpoint with minimal configuration. The codebase, benchmark queries, results, and demo recordings are available at \url{https://anonymous.4open.science/r/nl2sparql2-C171}. 
\end{itemize}
\section{Related Work}
\label{sec:related}



\textbf{GRASP}~\cite{walter2025grasp} is the prior work most similar to \kgv, employing an LLM agent with KG-agnostic tools for zero-shot SPARQL generation over arbitrary knowledge graphs. \kgv differs in three respects: the scale of required indexing, the inference of property semantics, and the retrieval mechanism. \emph{First}, GRASP indexes every entity, whereas \kgv indexes only classes, which scale with the ontology rather than the data. The class index is thus orders of magnitude smaller, as shown for two benchmark knowledge graphs (Table~\ref{tab:index_size}). The class index can also be built for far more OKN endpoints (Section~\ref{sec:feasibility}).\todo[color=cyan]{The overlap on GRASP is taking some extra space.}

\begin{table}[t]
\centering
\small
\setlength{\tabcolsep}{4pt}
\begin{tabular}{|l|r|r|r|r|}
\hline
 & \multicolumn{2}{c|}{\textbf{\kgv (classes)}} & \multicolumn{2}{c|}{\textbf{GRASP (entities)}} \\
\cline{2-3}\cline{4-5}
\textbf{KG} & \textbf{Count} & \textbf{Size} & \textbf{Count} & \textbf{Size} \\ \hline
DBLP & 35 & 2.7\,KB & 140.9M & 21\,GB \\
SOCKG & 91 & 5.3\,KB & 7.1M & 1.1\,GB \\ \hline
\end{tabular}
\caption{Indexed items and on-disk size of \kgv's class index vs.\ GRASP's entity index for the DBLP and SOCKG knowledge graphs.}
\label{tab:index_size}
\end{table}
\emph{Second}, GRASP differentiates properties using ontological domain/range annotations, which are not always present (the ontology-deficiency challenge discussed in Section~\ref{sec:introduction}); \kgv infers them directly from data via dedicated exploration tools. \emph{Third}, GRASP uses semantic and keyword search as separate signals across its tools, whereas \kgv combines them jointly via hybrid retrieval, which has been shown to outperform either signal alone in various applications~\cite{kuzi2020leveragingsemanticlexicalmatching,rayo2025hybridapproachinformationretrieval}.\todo[color=cyan]{The difference sounds pure engineering. We need to think hard about whether there is a way for us to claim more research contributions.}

\textbf{ToG}~\cite{sun2022think} pursues KG-agnostic question answering through an agentic LLM that traverses the knowledge graph hop-by-hop. Unlike \kgv, which explores latent structure and semantics to construct an executable SPARQL query, ToG traverses instance-level data node-by-node to locate answers. The two systems differ in three respects.  \emph{First}, in access assumptions: adapting ToG to a new knowledge graph requires downloading and locally deploying the entire graph and implementing a graph-specific Python traversal interface, which is infeasible in endpoint-only settings; \kgv uses a unified tool implementation that works across all knowledge graphs that follow the RDFS typing standard via standard SPARQL. \emph{Second}, in grounding: ToG assumes the presence of an initial set KG entities along with the question from which exploration will begin; \kgv explicitly treats class, property, and entity grounding as part of the endpoint-only task. \emph{Third}, in reasoning efficiency: ToG's search process grows proportionally with hop distance, the branching factor of intermediate neighborhoods, and the cardinality of candidate solution nodes, whereas \kgv's tool suite relies on SPARQL queries that can express traversals over arbitrarily many hops in a single execution while yielding results of any cardinality. Finally, ToG's evaluation covers only open-domain knowledge graphs, providing limited evidence of generalization to domain-specific knowledge graphs.

Other systems are tailored to a specific knowledge graph or domain.  \textbf{Spinach}~\cite{liu2024spinach} uses Wikidata-specific tools and an in-context learning agent to simulate expert exploration for question answering over Wikidata, while \textbf{Expasy}~\cite{emonet2024llm} presents a bioinformatics text-to-SPARQL system that mitigates hallucinations via schema validation, but relies on curated text--query pairs and KG-specific schemas. 
Their tight coupling to specific domains limits generalization to arbitrary knowledge graphs without substantial re-engineering, in contrast to the flexibility of \kgv's KG-agnostic architecture. 
\vspace{2mm}
\section{Methodology}
\label{sec:methodology}

\vspace{-2mm}
\kgv is a KG-agnostic agentic reasoning architecture equipped with dynamic exploration tools to discover latent graph structure, semantics, and instances to reflectively generate SPARQL queries from natural language questions. \kgv relies on the minimal assumption that the target knowledge graph adheres to W3C RDFS standards for class definitions---a convention adopted across almost all RDF knowledge graphs (Appendix~\ref{sec:appendix_rdf_typing}). Under this assumption, search and exploration tools use KG-agnostic SPARQL queries and run unchanged across KGs. 
\begin{figure*}[t!]
    \centering
    \makebox[\textwidth][c]{\includegraphics[width=1.05\textwidth, trim=0cm 0cm 0cm 0cm, clip]{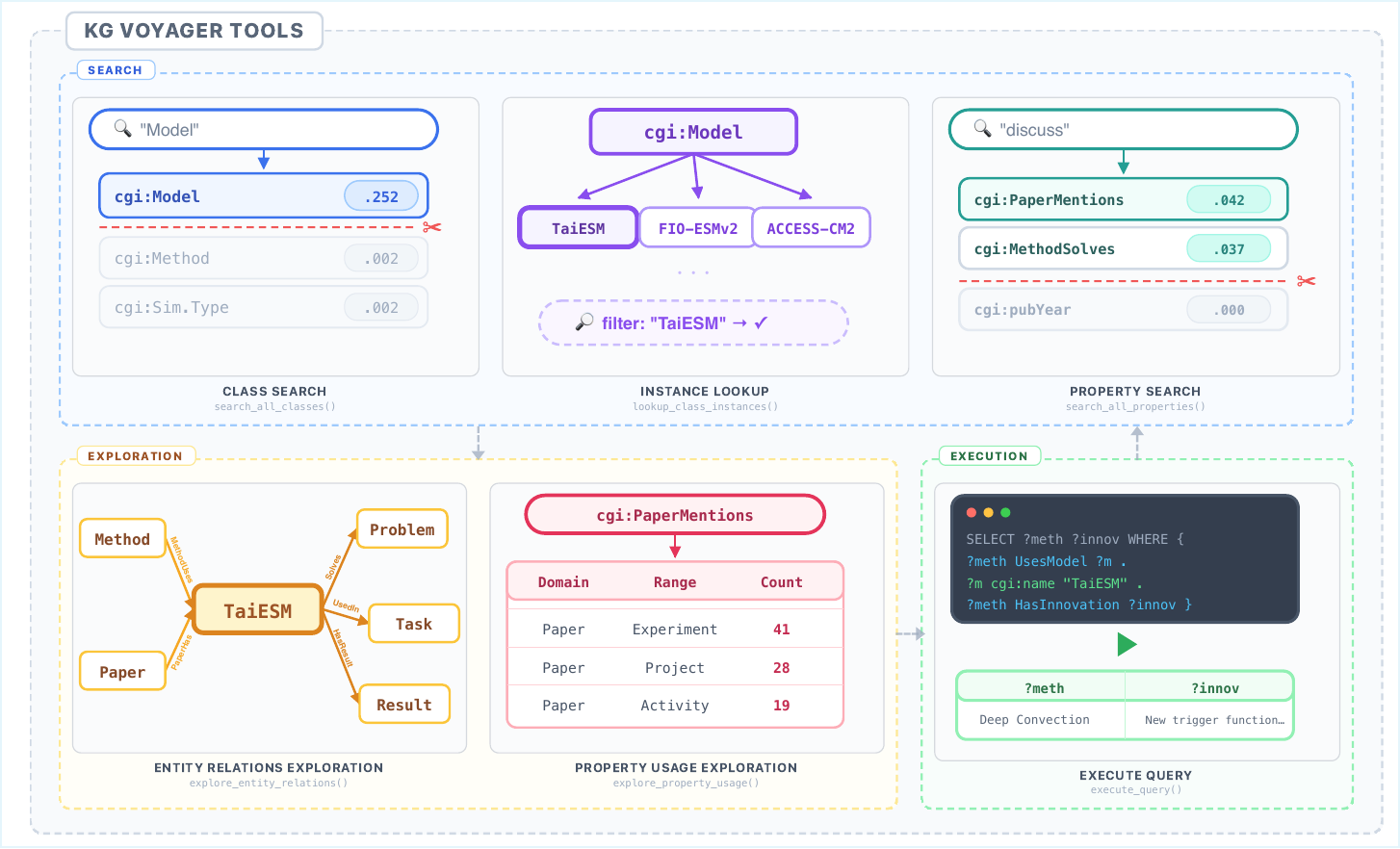}}
    \caption{The \kgv tool suite. Examples are simplified from runs on the Climate Models knowledge graph.} 
    \vspace{-3mm}
    \label{fig:kg_voyager_pipeline}
\end{figure*}

Given a natural language question, \kgv formulates SPARQL generation as an iterative exploration problem. The agent operates in a think-act-observe loop~\cite{yao2023react}, reasoning about what to do next, invoking a tool, and incorporating the result before deciding its next step. Its KG-agnostic tools are organized into three groups: (i)~\textit{search tools}, which retrieve candidate classes, node instances, and properties from natural language cues; (ii)~\textit{exploration tools}, which inspect how these candidates are used and connected in the knowledge graph; and (iii)~an \textit{execution tool}, which runs candidate SPARQL queries and returns feedback. The agent uses search to map question terms to classes, entities, and properties, each identified by an IRI. It then uses exploration to fill gaps in how these elements connect. Finally, it uses execution to validate and refine query drafts until it determines that a SPARQL query correctly answers the question. 

Figure~\ref{fig:kg_voyager_pipeline}\todo[color=cyan]{some of the fonts in the figure are really small. Are you able to increase them?} summarizes the tool suite. 
Search tools resolve classes, properties, and entities from natural-language cues. Exploration tools inspect entity neighborhoods and property usage to expose structure not immediately apparent from search results. \textit{execute\_query} validates partial or complete query drafts. The complete \kgv system prompt and tool specification are included in Appendix~\ref{sec:appendix_prompt_kgvoyager_tools}. We prompt the agent to generally follow a search, exploration, and execution sequence, mirroring the natural workflow of a human expert encountering an unfamiliar knowledge graph. Appendix~\ref{sec:appendix_tool_dynamics} confirms that models typically follow this ordering. The agent submits the final query through a dedicated \textit{answer} tool to mark task completion. Following GRASP ~\cite{walter2025grasp}, we also allow the agent to terminate early via a \textit{cancel} tool when generating a query is unfeasible for a given question under the knowledge graph. \todo[color=cyan]{It is just strange to use climate model to explain framework and DBLP for the end-to-end examples. And it is strange the example comes at the end. Maybe it is better to use the same concrete example and KG from the beginning.} 

\subsection{Search Tools}

\textbf{Class Search.}\hspace{1mm} The \textit{search\_all\_classes} tool accepts a natural language search term and returns a ranked list of classes whose IRIs or labels are semantically or lexically related to the term. Each result includes the class label, aliases, and instance count. On first invocation, the tool fetches a lightweight class index via SPARQL and caches it locally for all subsequent searches. To retrieve and rank results, we employ a hybrid approach that combines dense vector similarity with sparse BM25 lexical scoring  via reciprocal rank fusion~\cite{cormack2009reciprocal}, which has been shown to consistently outperform either signal alone in various applications ~\cite{kuzi2020leveragingsemanticlexicalmatching, rayo2025hybridapproachinformationretrieval}.

Each result is returned with a relevance score. To prevent low-quality candidates from polluting the agent's context, a phenomenon known as context rot~\cite{hong2025context}, we apply an adaptive filtering mechanism. This mechanism prunes candidates whose relevance score relative to the top-ranked result falls below a ratio $\delta$, while always retaining the top three candidates. We manually tune this hyperparameter and set $\delta=0.125$. 

\textbf{Instance Lookup.}\hspace{1mm}  The \textit{lookup\_class\_instances} tool retrieves instances of a given class, optionally filtered by an \textit{exact\_term}.
After identifying a candidate class IRI via \textit{search\_all\_classes}, the agent uses this tool to verify that the class contains the expected content or locate entities mentioned in the question.
The default lookup matches the \textit{exact\_term} against instance labels and IRIs. Searches using abbreviations, alternative spellings, or overly specific names may return no results because matching relies on lexical SPARQL string filters. To handle such cases, the tool supports a broader \textit{fallback\_term}\todo[color=cyan]{fallback term doesn't seem to be mentioned in tool specification} to retrieve a larger candidate set when the \textit{exact\_term} search fails. These candidates are then filtered by the semantic similarity of their labels to the \textit{exact\_term}. Separately, for knowledge graphs that store useful names or identifiers in literal-valued attributes rather than standard labels, the \textit{lookup\_class\_instances} tool also supports an \textit{all\_attributes} mode that further extends the search to all literal values of class instances, at the cost of slower execution. These class-scoped mechanisms enable precise entity resolution without an exhaustive entity index over the entire graph, which is infeasible for most endpoints.

\textbf{Property Search.}\hspace{1mm}  The \textit{search\_all\_properties} tool searches for properties using an approach similar to that used for class search. 
Given a search term, it returns a ranked list of property IRIs along with labels, aliases, usage counts as well as domains and ranges, if available. As with class search, the property index is fetched and cached on first use by enumerating all distinct predicates from the endpoint. Subsequent searches apply the hybrid retrieval and adaptive filtering methods used for class search to the locally cached property index.\todo[color=cyan]{Earlier the paper says KGVoyager only assumes class index. Isn't that it also needs property index?

\reply{
    GRASP also assumes/needs a property index but it would be more clunky and give reviewers more obvious avenues to attack if I mention property index in introduction

    In principle, KGVoyager can work without search all props (but not without search all classes) because of the explore entity relation tools which surfaces properties coming from entities. But it would take me some experiments to prove this

    Properties usually far less than entities and thus almost always could be queried efficiently. For example, in QueryBloom I query all DBPedia properties easily. I would need another study for it specific to KGVoyager but don't have time.
}} 

\subsection{Exploration Tools}

While the search tools resolve natural language terms to candidate class, property, or instance IRIs, they do not reveal how those IRIs are structurally embedded and connected in the graph. To understand the interaction of properties and the neighborhoods of entities, the agent requires tools that inspect the graph directly.

\textbf{Property Usage Exploration.}\hspace{1mm}  Many knowledge graphs lack explicit or accurate domain and range annotations, leading to ambiguity in property interpretation. The \textit{explore\_property\_usage} tool bridges this gap by examining how a property is actually used. Given a property IRI, it samples triples with that property and aggregates the subject and object classes involved, along with representative instances and their labels.  This allows the agent to verify that a candidate property empirically connects the kinds of entities expected by the question, before committing it to the query as well as to disambiguate properties with similar labels. For instance, while a property labeled ``located in'' connects organizations and companies to cities, a different property ``location'' may connect events to venues. 

\textbf{Entity Relation Exploration.}\hspace{1mm}  
The tool \textit{explore\_entity\_relations} inspects the local neighborhood of a given entity. From an entity IRI, it retrieves the incoming and outgoing properties along with example subjects or objects reached through each, including labels when available. This helps understand how the entity is embedded in the graph, particularly when property search has not surfaced the property needed for the question. Given a literal instead, the tool returns the entities with that value and the corresponding data properties.

\subsection{Execution Tools}

The \textit{execute\_query} tool runs a SPARQL query against the endpoint and returns the results. It serves a dual role: providing execution-based feedback from intermediate queries throughout the reasoning process, and executing the final query to obtain the answer. The agent can run intermediate queries to test hypotheses about graph structure, verify that queries produce non-empty results, and diagnose errors in partial queries. This iterative feedback is central to the agent's ability to self-correct and converge on accurate queries. 






\subsection{End-to-End Example}
\label{sec:end_to_end_example}
\begin{figure}[!htbp]
    \centering
    \includegraphics[width=\columnwidth, trim=0mm 10mm 0mm 0mm, clip]{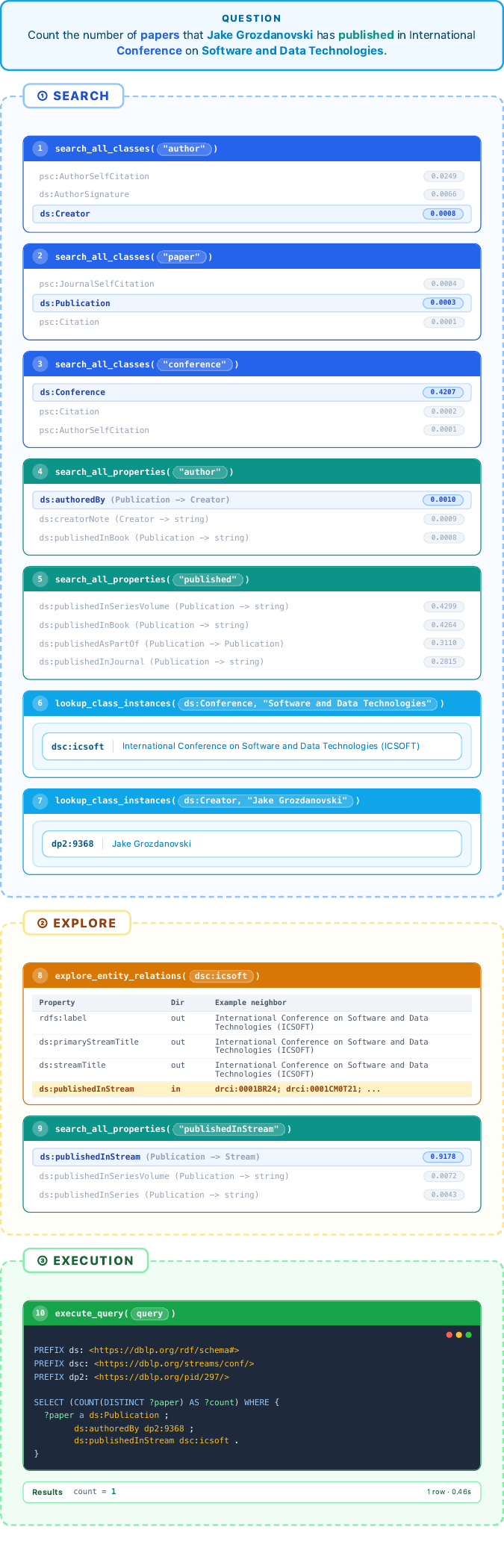}
    \caption{A complete \kgv example. Numbered blocks indicate tool calls in the order they were invoked by the agent. Abbreviations such as `ds:' denote namespace prefixes used in the underlying graph.}
    \vspace{-2mm}
    \label{fig:end_to_end_example}
\end{figure}
Figure~\ref{fig:end_to_end_example}\todo[color=cyan]{Currently Figure~\ref{fig:end_to_end_example} takes a whole column although there is some blank space. If you make the figure itself a bit more compact, we can avoid the blank. E.g., 1) Put the ``Question'' in the same line with the question itself; 2) reduce vertical gap below ``search'', ``explore'' and ``execution''. 3) reduce the margin beside the dashed bounding boxes.} shows \kgv resolving the question \textit{``Count the number of papers that Jake Grozdanovski has published in the International Conference on Software and Data Technologies''} in ten tool calls, over the DBLP knowledge graph.\todo[color=cyan]{``Explore'' in Figure~\ref{fig:end_to_end_example} should be ``Exploration'', since it uses ``Execution'' instead of ``Execute''. This will also be consistent with the other figure.} During \textbf{Search}, the agent decomposes the question into semantic search terms and resolves them to KG elements. 
The agent infers \textit{author}, implied by ``Jake Grozdanovski has published'' though not a literal query keyword, and maps it to \texttt{ds:Creator} (step~1), then maps \textit{paper} to \texttt{ds:Publication} (step~2) and \textit{conference} to \texttt{ds:Conference} (step~3). It reuses \textit{author} to surface the authorship property \texttt{ds:authoredBy} (step~4), while a search on \textit{``published''} (step~5) returns only journal/book properties, leaving the property linking publications to conference venues unresolved. The agent then grounds the two named entities: the conference name to \texttt{dsc:icsoft} (step~6) and the person name to \texttt{dp2:9368} (step~7). In the \textbf{Explore} phase, the agent inspects \texttt{dsc:icsoft} directly, which reveals the incoming relation \texttt{ds:publishedInStream}. A targeted follow-up search confirms it is the Publication~$\to$~Stream property required for the publication-conference relationship. In \textbf{Execution}, the agent has fully resolved the triple pattern \texttt{?paper ds:authoredBy dp2:9368; ds:publishedInStream dsc:icsoft}, executes the \textsc{count} query, and obtains the answer~\textbf{1}.
\section{Experiments}
\label{sec:evaluation}

\subsection{Feasibility Analysis}\label{sec:feasibility}

\begin{table}[t]
\centering
\setlength{\tabcolsep}{1pt}%
\small
\begin{tabular}{m{3.8cm}|cc|cc}
\hline
 & \multicolumn{2}{c|}{\textbf{Jena}} & \multicolumn{2}{c}{\textbf{QLever}} \\
\textbf{Knowledge Graph} & \textbf{Entity} & \textbf{Class} & \textbf{Entity} & \textbf{Class} \\ \hline
biobricks-aopwiki & $\checkmark$ & $\checkmark$ & $\times$ & $\checkmark$ \\
biobricks-ice & $\times$ & $\times$ & $\times$ & $\checkmark$ \\
biobricks-mesh & $\times$ & $\times$ & $\times$ & $\checkmark$ \\
biobricks-pubchem-annotations & $\times$ & $\checkmark$ & $\times$ & $\checkmark$ \\
biobricks-tox21 & $\checkmark$ & $\checkmark$ & $\times$ & $\checkmark$ \\
biobricks-toxcast & $\times$ & $\times$ & $\times$ & $\checkmark$ \\
biohealth & $\times$ & $\checkmark$ & $\times$ & $\checkmark$ \\
climatemodelskg & $\checkmark$ & $\checkmark$ & $\times$ & $\checkmark$ \\
dream-kg & $\checkmark$ & $\checkmark$ & $\times$ & $\checkmark$ \\
fio-kg & $\times$ & $\times$ & $\times$ & $\checkmark$ \\
gene-expression-atlas-okn & $\times$ & $\checkmark$ & $\times$ & $\checkmark$ \\
geoconnex & $\times$ & $\checkmark$ & $\times$ & $\checkmark$ \\
hydrology-kg & {\small\faClock[regular]} & $\checkmark$ & $\times$ & $\checkmark$ \\
nasa-gesdisc-kg & $\times$ & $\checkmark$ & $\times$ & $\checkmark$ \\
nde & $\times$ & $\checkmark$ & $\times$ & $\checkmark$ \\
neighborhood-information-kg & $\checkmark$ & $\checkmark$ & $\times$ & $\checkmark$ \\
rural-kg & $\checkmark$ & $\checkmark$ & $\times$ & $\checkmark$ \\
sawgraph & $\times$ & $\times$ & $\times$ & $\checkmark$ \\
scales & $\times$ & $\checkmark$ & $\times$ & $\checkmark$ \\
secure-chain-kg & $\times$ & $\checkmark$ & $\times$ & $\checkmark$ \\
semopenalex & $\times$ & $\times$ & $\times$ & $\checkmark$ \\
sockg & $\times$ & $\checkmark$ & $\times$ & $\checkmark$ \\
spatial-kg & {\small\faClock[regular]} & {\small\faClock[regular]} & $\times$ & $\checkmark$ \\
spoke-genelab & $\times$ & $\checkmark$ & $\times$ & $\checkmark$ \\
sud-okn & $\times$ & $\times$ & $\times$ & $\checkmark$ \\
ubergraph & $\times$ & $\checkmark$ & $\times$ & $\checkmark$ \\
uf-okn & {\small\faClock[regular]} & $\checkmark$ & $\times$ & $\checkmark$ \\
wikidata & $\times$ & $\times$ & $\times$ & {\small\faClock[regular]} \\ \hline
\textbf{Success Rate} & \textbf{21.4\%} & \textbf{67.9\%} & \textbf{0\%} & \textbf{96.4\%} \\ \hline
\end{tabular}
\caption{Indexing feasibility ($\checkmark$: Success, $\times$: Failed, {\small\faClock[regular]}: Timeout) across 28 KG SPARQL endpoints from \texttt{okn.us} under two settings.}
\label{tab:feasibility_comparison}
\end{table}

We assessed the feasibility of GRASP's entity index and \kgv's class index by attempting to build each over 28 KG SPARQL endpoints from \texttt{okn.us}~\cite{frink2026}, a catalog of graphs developed under the U.S.~NSF Proto-OKN initiative. For GRASP, we used its official entity index query. 
We compared two settings: (i) the individual Apache Jena endpoints of \texttt{okn.us} and (ii) its unified federated QLever endpoint~\cite{Bast2017QLever}, which \texttt{okn.us} plans to adopt in place of Jena for performance reasons.\todo[color=cyan]{What's the current state of this? Are we able to write more accurately?\reply{They likely transitioned already, but I decided not to repeat the analysis because I was afraid these percentages could change a lot given their substantial changes in deployment. What's here isn't the most up to date but reflects a deployment scenario that existed during paper writing.}} 

Table~\ref{tab:feasibility_comparison} reports the results. Under Jena, class indexing succeeded for $\sim$68\% of endpoints while entity indexing succeeded for only $\sim$21\%, confirming the impracticality of GRASP's entity-index requirement; the Jena entity-indexing failures returned opaque server errors with no actionable diagnostic information. The gap widens under QLever: class indexing succeeded for $\sim$96\% of endpoints (27/28), while entity indexing failed on all 28, with failures manifesting as server memory crashes—an interpretable outcome given the size of entity indices. These results show that the feasibility of entity-level indexing is highly sensitive to the underlying SPARQL engine and knowledge graph. 

\begin{table*}[!htbp]
\centering
\setlength{\tabcolsep}{2pt}%
\scriptsize
\begin{tabular}{l|rrr|rrr|rrr|rrr}
\hline
 & \multicolumn{3}{c|}{\textbf{F1 (\%)}} & \multicolumn{3}{c|}{\textbf{Cost (\$)}} & \multicolumn{3}{c|}{\textbf{Iterations}} & \multicolumn{3}{c}{\textbf{Time (s)}} \\
\textbf{Benchmark} & \sbkgv & \textbf{GRASP} & \textbf{$\Delta$} & \sbkgv & \textbf{GRASP} & \textbf{$\Delta$} & \sbkgv & \textbf{GRASP} & \textbf{$\Delta$} & \sbkgv & \textbf{GRASP} & \textbf{$\Delta$} \\ \hline
Climate Models KG & \textbf{69.8} & 67.2 & +2.6 & \textbf{2.66} & 4.57 & \textminus 41.8\% & \textbf{19.6} & 23.1 & \textminus 3.5 & \textbf{75.1} & 119.7 & \textminus 37.3\% \\
SOCKG & \textbf{71.3} & 68.9 & +2.4 & \textbf{2.20} & 4.00 & \textminus 45.0\% & \textbf{19.3} & 25.0 & \textminus 5.7 & \textbf{57.0} & 108.6 & \textminus 47.5\% \\
DREAM-KG & \textbf{49.7} & 46.9 & +2.8 & \textbf{3.06} & 5.06 & \textminus 39.5\% & \textbf{24.2} & 30.5 & \textminus 6.3 & \textbf{55.9} & 92.8 & \textminus 39.8\% \\
DBLP & \textbf{67.5} & 60.8 & +6.7 & \textbf{1.79} & 2.13 & \textminus 16.0\% & 15.1 & 15.1 & 0.0 & 154.2 & \textbf{128.7} & +19.8\% \\ \hline
\textbf{Average} & \textbf{64.6} & 60.9 & +3.7 & \textbf{2.43} & 3.94 & \textminus 38.3\% & \textbf{19.6} & 23.4 & \textminus 3.8 & \textbf{85.6} & 112.4 & \textminus 23.8\%\\ \hline
\end{tabular}
\caption{Performance by benchmark, averaged across four backbone models. GRASP is evaluated under its best-case setting (entity index available).}
\label{tab:benchmark_avg_performance}
\end{table*}

\begin{figure*}[!htbp]
\centering
\includegraphics[width=\textwidth]{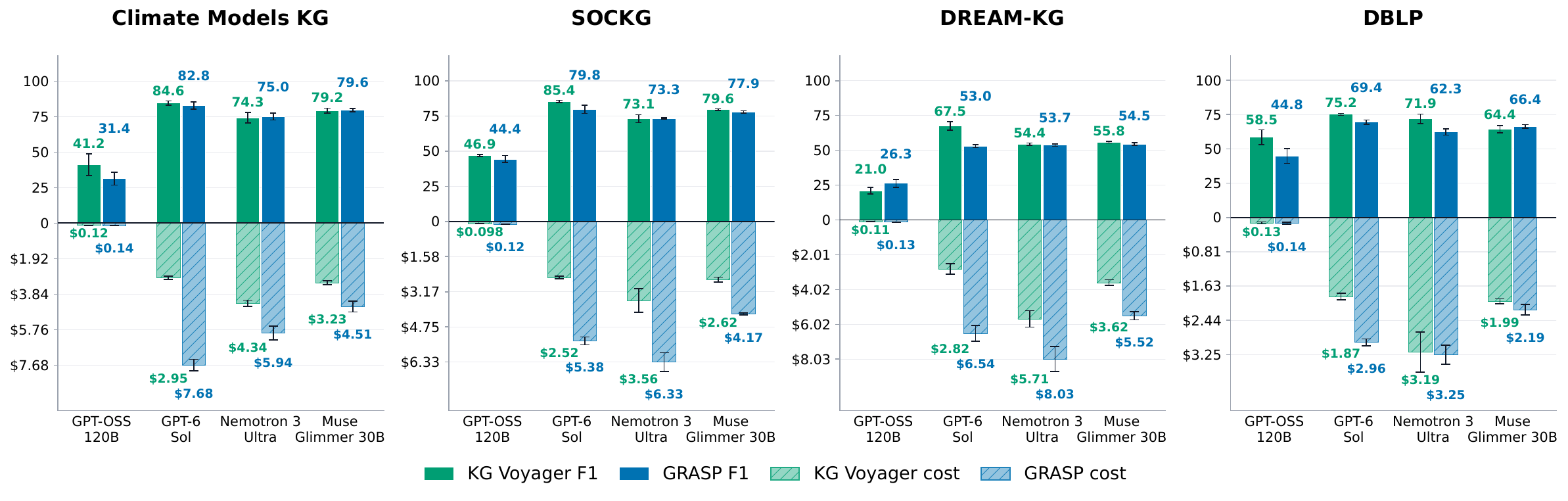}
\caption{Per-model F1 and monetary cost by benchmark. F1 bars extend above the baseline, cost (hatched) below; error bars indicate run-to-run standard deviation across three independent runs.}
\vspace{-2mm}
\label{fig:benchmark_cost_f1_mirror_bars}
\end{figure*}

\subsection{Benchmarks and Models}
\label{sec:benchmarks}
We evaluated question answering on four domain-specific benchmarks: Climate Models KG~\cite{adamu2025climatepub4kg} (108 questions on climate-science literature), SOCKG~\cite{sockg2025} (98 questions on soil-carbon modeling), DREAM-KG (88 questions on local government service data), and DBLP-QuAD (100 questions on scholarly metadata). The first three are associated with \texttt{okn.us}~\cite{frink2026}, with question--SPARQL pairs supplied by the respective KG developers. DBLP-QuAD is a public question--SPARQL benchmark; we randomly sampled 100 questions from its test split~\cite{dblpquad_hf}. For brevity, we refer to each benchmark by the name of its knowledge graph. The question--SPARQL pairs used in our evaluation are available through our repository and in the supplementary data file accompanying this manuscript. 

Although Section~\ref{sec:feasibility} shows that constructing GRASP's entity index from endpoint access is infeasible for most knowledge graphs, we evaluate GRASP under its \emph{best-case} conditions by providing the required entity index. We obtained this index through one of three routes: (i) building it directly on relatively small KGs whose Jena endpoint can serve the entity-index query, as for Climate Models KG and DREAM-KG; (ii) locally redeploying the knowledge graph and extracting the index from the deployment, as for SOCKG, whose \texttt{okn.us} endpoint fails entity index creation (Section~\ref{sec:feasibility}); or (iii) using a pre-built index, as provided by GRASP's authors for DBLP-QuAD. 

\kgv assumes only endpoint access. For SOCKG, the KG's authors provided the ontology as a Turtle file, which we use to construct an additional Grep Ontology baseline. This baseline is built on the Unix \textit{grep} command, which coding agents commonly use to search code. It uses the same backbone models with only two tools: \textit{grep} over the ontology file and SPARQL query execution. \kgv uses its standard tool suite and is not given the ontology file. This comparison tests whether \kgv, using only endpoint access, can outperform a baseline given the full ontology.

We ran our experiments on these benchmarks with four backbone models spanning closed and open systems of varying size and agentic capability. 
GPT-6 Sol (OpenAI, 2026) is a cost-efficient closed model built for complex coding and agentic workflows ~\cite{openai2026gpt6sol}. GPT-OSS-120B (OpenAI, 2025) is an open-weight mixture-of-experts reasoning model with 117B total and 5.1B active parameters, built with tool calling support~\cite{openai2025gptoss}. Nemotron 3 Ultra (NVIDIA, 2026) is an open hybrid Mamba--Transformer mixture-of-experts model with 550B total and 55B active parameters, targeting reasoning and agentic workflows~\cite{nvidia2026nemotron3ultra}. Muse Glimmer 30B (Meta, 2026) is an open 30B-parameter dense model distilled from Muse Spark for local agentic tasks with failure recovery~\cite{meta2026museglimmer}. The three open models were accessed via OpenRouter. Full implementation hyperparameters are in Appendix~\ref{sec:appendix_implementation}.

\subsection{QA Performance}\label{sec:qa_performance}

We compared \kgv against the prior state of the art, GRASP~\cite{walter2025grasp}, on the benchmarks and models above. For each combination of method, benchmark and model, we average three independent runs and report F1, monetary cost, agent iterations, and wall-clock time. Following GRASP, we adopt the row-major F1 metric of Spinach~\cite{liu2024spinach}.\todo{would any comparison make t-test or z-test meaningful? We might want to include some statistical test results, so that the reviewers will not pick on this. \reply{Not straightforward given the remaining time}}

Table~\ref{tab:benchmark_avg_performance} reports results averaged across runs and models; Figure~\ref{fig:benchmark_cost_f1_mirror_bars} shows per-model results. \kgv improves F1 on all four benchmarks, with gains ranging from 2.4 percentage points on SOCKG to 6.7 points on DBLP. It also reduces cost on every benchmark and requires fewer agent iterations on all benchmarks except DBLP, where the two systems tie. Runtime is lower by 37--48\% on three benchmarks; DBLP is the exception (19.8\% higher) because its template-based questions focus heavily on specific entity instances (e.g., paper and author names), which GRASP resolves with a single lookup in its pre-supplied entity index, whereas \kgv must first identify the relevant class and then search within it.\todo{The table shows GRASP and KGVoyager use the same iteration count on DBLP; contradicts ``a single look-up''.\reply{I can explain more deeply in the paper but it would be confusing. GRASP has an entity index, so it can find entities instantly. \kgv can find the class via its class index, but then needs to wait for the instance lookup tool, which takes time as it searches with a SPARQL query within the class instances, with no cache.}} Overall, \kgv also makes fewer tool calls and SPARQL executions per question (Appendix~\ref{sec:appendix_tool_invocations}). We also verify that the benchmarks are nontrivial and analyze performance by query complexity (Appendix~\ref{sec:appendix_complexity}). Finally, we categorize \kgv's failure cases in a detailed failure analysis (Appendix~\ref{sec:appendix_error_analysis}).

\begin{figure}[t]
\centering
\includegraphics[width=0.92\columnwidth]{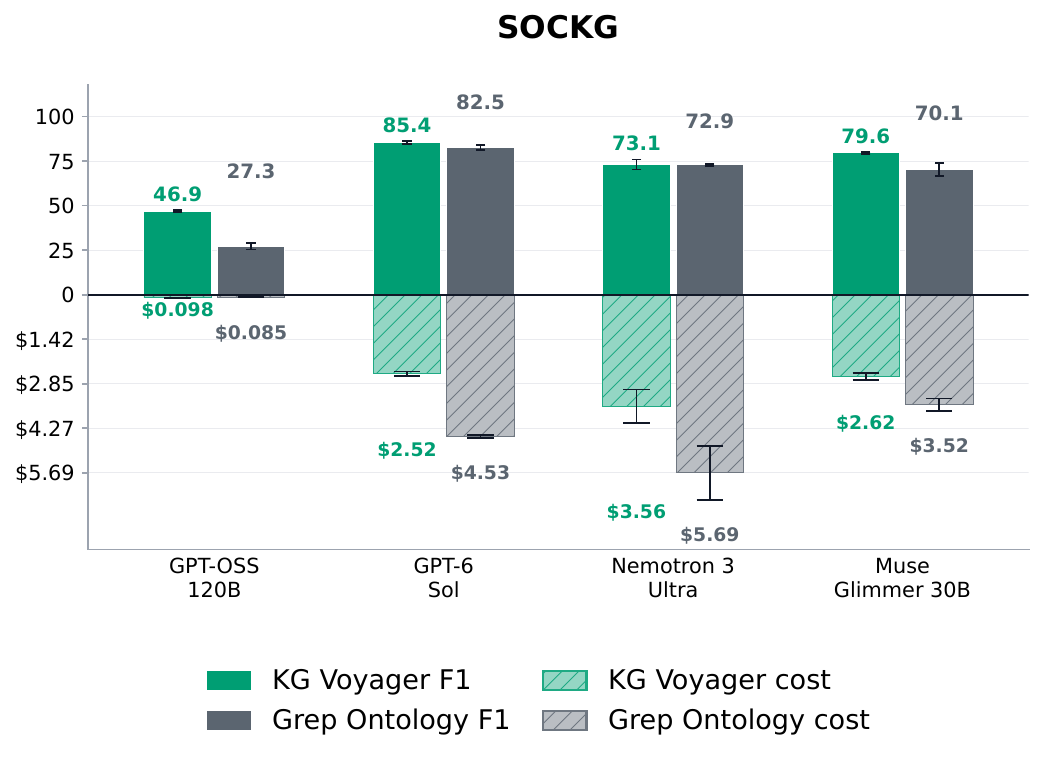}
\caption{Per-model F1 (top) and monetary cost (hatched, bottom) of \kgv and the Grep Ontology baseline on SOCKG; error bars show standard deviation over three runs.}
\label{fig:kgvoyager_grep_f1_cost}
\end{figure}

Beyond GRASP, we compared \kgv against the Grep Ontology baseline on SOCKG (Section~\ref{sec:benchmarks}). The results in Figure~\ref{fig:kgvoyager_grep_f1_cost} show that, despite having direct access to the ontology, the baseline trails \kgv by 8.1 points of F1 on average (63.2\% vs.\ 71.3\%), while \kgv also reduces cost by 36.4\%. We also observed an 18.7\% decrease in runtime. The largest F1 gains are observed for the two smaller models, GPT-OSS-120B (+19.6 points) and Muse Glimmer 30B (+9.5). We attribute these differences to context rot~\cite{hong2025context}, as repeated grep calls return long outputs that accumulate in the context, which raises token cost and tends to degrade smaller models more.

\subsection{Ablation Study}\label{sec:ablation}

We ablate three components of \kgv: entity-relation and property-usage exploration, which we remove, and hybrid retrieval, which we replace with keyword-only retrieval in class and property search. Search and execution tools are always retained, as an endpoint-only system cannot ground terms or test queries without them. Each component is evaluated on at least two models and two benchmarks.

\begin{table}[t]
\centering
\setlength{\tabcolsep}{3pt}%
\renewcommand{\arraystretch}{1.3}%
\scriptsize
\begin{tabular*}{\columnwidth}{@{}l@{\extracolsep{\fill}}rrrrrr@{}}
\hline
 & \multicolumn{2}{c}{\textbf{F1 (\%)}} & \multicolumn{2}{c}{\textbf{Cost (\$)}} & \multicolumn{2}{c}{\textbf{Iter.}} \\
\textbf{Model, benchmark} & Full & Abl. & Full & Abl. & Full & Abl. \\ \hline
\multicolumn{7}{l}{\textbf{Entity-relation exploration} (w/o \textit{explore\_entity\_relations})} \\
GPT-OSS, SOCKG & \textbf{46.9$\pm$.6} & 38.3$\pm$1.1 & \textbf{.098} & .098 & 15.9 & \textbf{15.8} \\
GPT-6 Sol, SOCKG & \textbf{85.4$\pm$.8} & 83.9$\pm$.6 & \textbf{2.52} & 2.74 & \textbf{10.7} & 11.4 \\
GPT-OSS, DBLP & \textbf{58.5$\pm$5.3} & 55.1$\pm$1.3 & .128 & \textbf{.081} & \textbf{10.8} & 12.5 \\
\hline
\multicolumn{7}{l}{\textbf{Property-usage exploration} (w/o \textit{explore\_property\_usage})} \\
Nemotron 3, SOCKG & \textbf{75.5$\pm$2.0} & 74.8$\pm$3.3 & \textbf{4.46} & 5.09 & \textbf{20.9} & 22.4 \\
GPT-OSS, Climate & \textbf{33.9$\pm$3.8} & 21.4$\pm$3.7 & .116 & \textbf{.094} & 16.6 & \textbf{14.7} \\
Muse 30B, SOCKG & \textbf{79.9$\pm$1.0} & 75.7$\pm$1.3 & \textbf{2.72} & 2.94 & \textbf{28.5} & 31.3 \\
\hline
\multicolumn{7}{l}{\textbf{Hybrid retrieval} (keyword-only instead of hybrid)} \\
GPT-OSS, Climate & \textbf{41.2$\pm$7.7} & 36.3$\pm$2.1 & .122 & \textbf{.115} & \textbf{16.1} & 18.6 \\
GPT-OSS, SOCKG & \textbf{46.9$\pm$.6} & 20.4$\pm$2.3 & .098 & \textbf{.094} & \textbf{15.9} & 19.7 \\
Muse 30B, Climate & \textbf{79.2$\pm$1.8} & 78.1$\pm$.9 & \textbf{3.23} & 3.86 & \textbf{30.7} & 33.9 \\
GPT-6 Sol, DREAM & \textbf{67.5$\pm$3.0} & 65.7$\pm$1.2 & \textbf{2.82} & 3.13 & \textbf{12.8} & 13.6 \\
\hline
\end{tabular*}
\caption{Ablation study (Abl.\ = component removed). F1 is mean $\pm$ std.\ over three independent runs; cost and iterations are means.}
\label{tab:ablation}
\end{table}

Table~\ref{tab:ablation} reports the ablation results. Removing \textit{explore\_entity\_relations} reduces F1 in all three pairs (by 1.5 to 8.6 points). 
Removing \textit{explore\_property\_usage} lowers F1 by 12.5 points on Climate Models KG with GPT-OSS-120B and 4.2 points on SOCKG with Muse Glimmer 30B, 
while the 0.7-point drop for Nemotron 3 Ultra on SOCKG falls within run-to-run variability. 
Only GPT-OSS-120B on Climate Models KG pays more cost and iterations with the tool; for the other two pairs, the tool lowers both. Both the full and ablated variants hide the property domain/range search tool, because, per the ontology deficiency challenge (Section~\ref{sec:introduction}), we assume such annotations are generally absent, even though some of these KGs may have them. This shows that property-usage exploration recovers useful semantic information directly from graph data when ontological annotations are unavailable.
Finally, hybrid retrieval improves F1 over keyword-only search in all four pairs (+1.1 to +26.5 points) and reduces iterations in all four, showing that combining semantic and lexical signals improves search robustness.
\section{Conclusion}

We presented \kgv, a KG-agnostic framework that answers natural language questions over arbitrary RDF knowledge graphs given only SPARQL endpoints. Our feasibility study shows that \kgv's lightweight class index succeeds on $\sim$96\% of \texttt{okn.us} endpoints under QLever, compared to 0\% for the entity index required by GRASP. Across four domain-specific benchmarks and four backbone models, \kgv improves F1 over GRASP while generally reducing cost, runtime, and iterations, showing that endpoint-only KGQA is viable where curated resources are unavailable. 
\section*{Limitations}

Although \kgv requires no KG-specific ontology, text--query examples, or graph-specific tools and assumes that the knowledge graph follows the RDF standard for defining types (\texttt{rdf:type}), which is broadly adopted in practice (Appendix~\ref{sec:appendix_rdf_typing}), knowledge graphs that use their own typing conventions, such as Wikidata, are not directly supported. Supporting them would require adapter tools, which we leave for future work.

A second limitation is data availability: domain-specific text--SPARQL benchmarks remain scarce, and despite reaching out to many KG authors and maintainers, we could not gather larger datasets. More public domain-specific KGQA benchmarks are needed and we envision that automated benchmark generation could be promising.
\section*{Ethical Considerations}

All knowledge graphs endpoints and benchmarks used in this work are publicly available (see links in our repository~\footnote{\url{https://anonymous.4open.science/r/nl2sparql2-C171}}), and no private or personal data was collected or processed in any part of our experiments. We use existing artifacts only for their intended research and benchmarking purposes. Released artifacts are intended for research, evaluation, and reproducibility, with third-party resources remaining subject to their original terms. 

Users of \kgv should be aware that, like any LLM-based system, it may generate syntactically valid SPARQL queries that are nonetheless semantically incorrect. In domains where query results inform decisions, such as scientific or policy contexts, outputs should be verified rather than trusted uncritically.

Finally, LLM inference carries an environmental cost. We note, however, that \kgv reduces monetary cost by ${\sim}$38\% and wall-clock runtime by ${\sim}$24\% relative to the prior state-of-the-art system, partially mitigating this concern. 

\bibliography{references}

\vspace{5mm}

\appendix
\section*{Appendix}

\noindent The appendix is organized as follows. Appendix~\ref{sec:appendix_implementation} lists implementation details and hyperparameters. Appendix~\ref{sec:appendix_rdf_typing} measures how widely \texttt{rdf:type} is used across \texttt{okn.us} endpoints, supporting \kgv's RDF typing assumption. Appendix~\ref{sec:appendix_complexity} analyzes performance as a function of query complexity. Appendix~\ref{sec:appendix_tool_dynamics} examines when the agent uses each tool group, and Appendix~\ref{sec:appendix_tool_invocations} reports per-question tool-call and SPARQL-execution counts. Appendix~\ref{sec:appendix_error_analysis} presents a failure analysis of \kgv, and Appendix~\ref{sec:appendix_prompt_kgvoyager_tools} gives the full system prompt and tool specification.

\section{Implementation Details}
\label{sec:appendix_implementation}

Table~\ref{tab:implementation_details} lists the key implementation parameters used across all experiments.

\vspace{2mm}
\begin{table}[!htbp]
\centering
\setlength{\tabcolsep}{4pt}%
\small
\begin{tabular}{p{2.5cm}l}
\hline
\textbf{Parameter} & \textbf{Value} \\ \hline
Embedding model & \texttt{BAAI/bge-small-en-v1.5} \\
Lexical retrieval & BM25 top-$k$: 20 \\
Dense retrieval & Dense top-$k$: 10 \\
Fusion & RRF, equal weights, rank constant 61 \\
Re-ranker & \texttt{BAAI/bge-reranker-base}, top-$k$: 10 \\
Adaptive filtering $\delta$ & 0.125, minimum 3 rows retained \\
Property-usage\newline sample size & 3000 triples \\
Vector index & In-memory, cosine similarity \\
Caching & Embeddings stored on disk; reused \\
 & unless metadata changes \\ \hline
\end{tabular}
\vspace{1mm}
\caption{Implementation parameters.}
\label{tab:implementation_details}
\end{table}
\section{Prevalence of RDFS Typing}
\label{sec:appendix_rdf_typing}

\kgv makes only one structural assumption: instances are assigned to classes via the W3C RDFS predicate \texttt{rdf:type}~\cite{w3c_rdf_schema}. Its class index and tools use this predicate to discover classes. To check how practical this assumption is, we counted typing triples and all triples via a SPARQL query on the 28 \texttt{okn.us} knowledge graphs from our feasibility study (Section~\ref{sec:feasibility}). The \texttt{wikidata} endpoint returned an error page instead of query results; however, Wikidata is known not to use \texttt{rdf:type} for typing entities, as it relies on its own type system.

\begin{figure}[!htbp]
    \centering
    \includegraphics[width=\columnwidth]{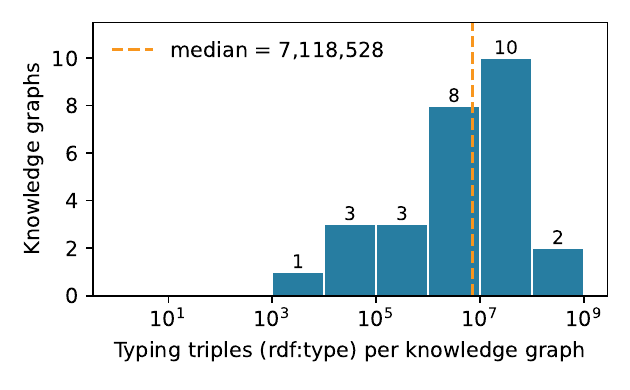}
    \caption{Number of \texttt{rdf:type} triples per \texttt{okn.us} knowledge graph (27 graphs that returned a count; bins are powers of ten).}
    \label{fig:rdf_typing}
\end{figure}

We consider a graph to follow the RDFS typing standard if it contains at least 1,000 typing triples. Of the 28 graphs, 27 (${\sim}$96\%) meet this criterion (median of ${\sim}$7.1 million typing triples over the 27 graphs that returned a count; Figure~\ref{fig:rdf_typing}), suggesting that the assumption is reasonable for most knowledge graphs. The only exception is \texttt{wikidata}. This scan is unlikely to be biased by platform policy, since \texttt{okn.us} only asks contributors for registry metadata and does not prescribe how entities are typed~\cite{frink2026}. The prevalence of \texttt{rdf:type} instead reflects its status as the W3C-recommended standard for assigning resources to classes~\cite{w3c_rdf_schema}.
\section{Query Complexity Analysis}
\label{sec:appendix_complexity}

To analyze performance as a function of question complexity and to ensure the benchmarks used for evaluation are nontrivial, we parsed the gold SPARQL queries corresponding to the benchmark questions and measured the complexity of the underlying query subgraph by counting triple patterns in the translated SPARQL algebra, together with the number of distinct special SPARQL constructs (aggregation, \texttt{FILTER}, \texttt{OPTIONAL}, \texttt{UNION}, \texttt{MINUS}, \texttt{EXISTS}, subqueries, and property paths). We computed paired F1 differences (\kgv minus GRASP) for each question, averaging the three runs within each backbone model and then averaging across models.

\begin{figure}[!htbp]
    \centering
    \includegraphics[width=\columnwidth]{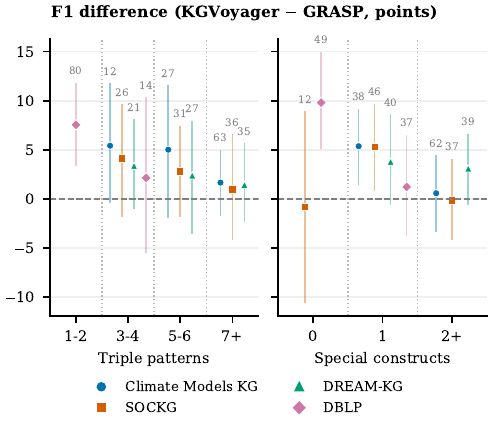}
    \caption{Mean paired F1 difference (\kgv minus GRASP) per benchmark by number of triple patterns (left) and distinct special SPARQL constructs (right) in the gold query. Points show means with 95\% bootstrap confidence intervals over questions; numbers above the intervals are the number of questions; bins with fewer than 10 questions are omitted.}
    \label{fig:complexity_analysis}\vspace{3mm}
\end{figure}

The benchmarks differ in the complexity of their gold queries, measured by the number of triple patterns, which reflects the size of the subgraph matched by the SPARQL query, and by the number of special SPARQL constructs, as shown in Figure~\ref{fig:complexity_analysis}.\todo[color=cyan]{This Figure looks a bit overly crowded; also need larger gaps between subfirue caption and legends.\reply{Gap fixed. As for the figure being crowded, I tried many alternatives while making it, such as a pairwise grid and stacking one panel on top of another, but they were worse: the first because of high standard deviations, the second because of different axes.}} DBLP contains simpler queries than other benchmarks, with at most four triple patterns, whereas the other benchmarks mostly contain queries with five or more. We observe that \kgv's improvement over GRASP drops for the most complex queries, on which both systems perform worse; for example, pooled F1 is 58.8\% for \kgv and 57.4\% for GRASP on queries with seven or more triple patterns, compared with 66.7--70.8\% and 62.9--67.4\% on queries with three to six.
\section{Tool-Use Dynamics}
\label{sec:appendix_tool_dynamics}

We analyze whether the agent follows the search--explore--execute progression described in Section~\ref{sec:methodology}. For each tool call, we compute its relative step index, defined as position divided by total run iterations, and pool values across the three \kgv runs per benchmark-model pair.

\begin{figure}[t]
\centering
\includegraphics[width=\columnwidth]{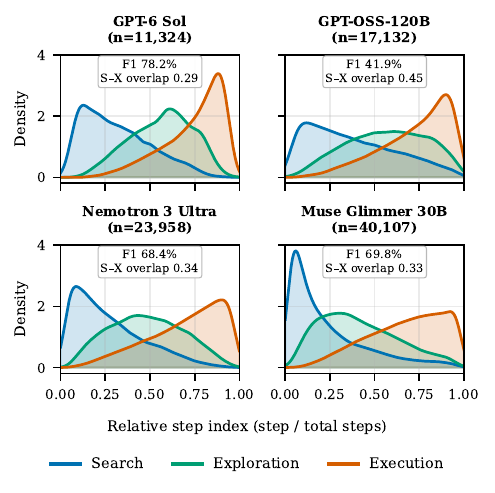}
\vspace{1mm}
\caption{Relative timing of tool use by model: kernel density estimates of the relative step index for the search, exploration, and execution tool groups. Each panel reports the model's \kgv F1 averaged over the four benchmarks and the S--X overlap, the overlap coefficient between the search and execution densities (lower means cleaner phase separation).}
\label{fig:tool_relative_index_distributions}\vspace{2mm}
\end{figure}

Figure~\ref{fig:tool_relative_index_distributions} shows all four models. All follow the intended temporal structure on average: search early, exploration in the middle, and execution later. GPT-6 Sol, the best-performing model (78.2\% F1), shows the cleanest phase separation, while GPT-OSS-120B, the weakest (41.9\% F1), shows the most overlap. The other two models have similar F1 (68.4\% and 69.8\% for Nemotron 3 Ultra and Muse Glimmer 30B) and similar overlap (0.34 and 0.33). This shows that phase mixing can provide some signal about model performance.
\section{Tool Invocation Distributions}
\label{sec:appendix_tool_invocations}

\begin{figure}[t]
    \centering
    \includegraphics[width=\columnwidth]{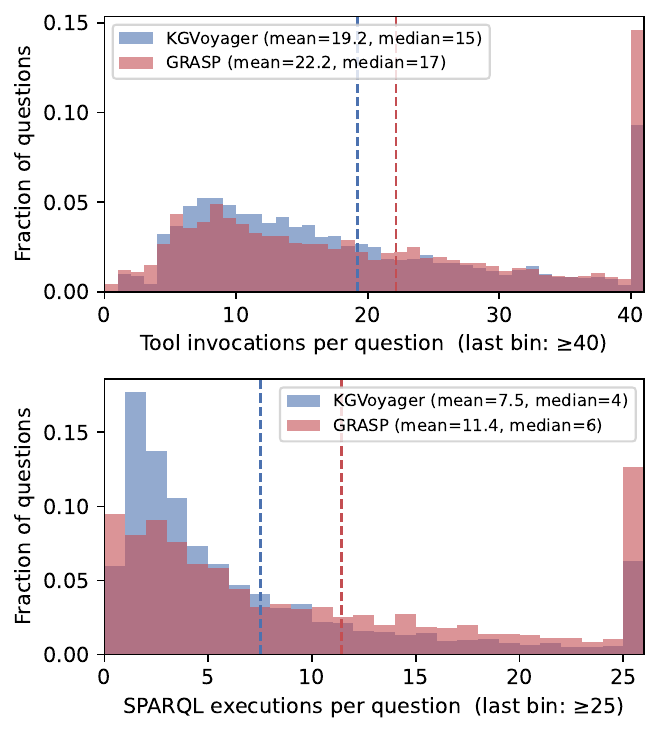}
    \vspace{1mm}
    \caption{Distribution of tool invocations (top) and explicit SPARQL executions (bottom) per question. Dashed lines indicate means.}
    \label{fig:tool_invocations}\vspace{2mm}
\end{figure}

Figure~\ref{fig:tool_invocations} reports the distribution of tool invocations per question (top) and the number of explicit SPARQL executions per question (bottom). Both analyses pool agent executions across all questions in the four benchmarks, four backbone models, and three runs. The distributions show that \kgv requires fewer tool invocations (19.2 vs.\ 22.2 on average) and SPARQL executions (7.5 vs.\ 11.4) per question than GRASP, which indicates a more efficient interaction with the knowledge graph.

\section{Failure Analysis}
\label{sec:appendix_error_analysis}

To understand the scenarios in which \kgv breaks down, we analyze complete failures: (model, KG, question) triples that scored an F1 of 0 across all three independent runs. Considering all the valid gold queries, this criterion yields 156 complete failures across the four models and benchmarks. We attribute each failure to a primary cause using a two-level taxonomy of five coarse-grained error groups, each split into fine-grained categories.

Table~\ref{tab:error_groups} summarizes the coarse-grained distribution. Tables~\ref{tab:error_concept}--\ref{tab:error_other} detail the fine-grained categories with representative examples.

The largest group, query generation/execution (36\%), consists almost entirely of early stopping, as seen in Table~\ref{tab:error_execution}, mostly by smaller models such as GPT-OSS-120B that appear to struggle with long contexts. Another 31\% represent genuine reasoning errors (concept resolution and query construction). As seen in Tables~\ref{tab:error_concept} and~\ref{tab:error_construction}, these are spread evenly across wrong classes, properties, and entities, and across missing constraints and incorrect joins. Answer representation (18\%; Table~\ref{tab:error_repr}) covers answers whose content is analogous to that of the gold query but whose form or shape differs. For example, returning paper IDs instead of paper names or using a SELECT query instead of an ASK query for a yes/no question.

\begin{table}[!htbp]
\centering
\setlength{\tabcolsep}{3pt}%
\renewcommand{\arraystretch}{1.3}%
\small
\vspace{2mm}
\begin{tabular}{p{1.8cm}p{4.2cm}rr}
\hline
\textbf{Group} & \textbf{Definition} & \textbf{$n$} & \textbf{\%} \\ \hline
Concept\newline Resolution & Agent linked a question concept to the wrong instance, class, or predicate IRI & 29 & 19 \\ \hline
Query\newline Construction & Correct terms, but wrong constraints, relation direction, or join path & 20 & 13 \\ \hline
Answer\newline Representation & Semantically analogous answer scored wrong due to not matching result form or shape & 28 & 18 \\ \hline
Query Gen.\newline / Execution & No usable final query was produced or a well-formed query failed to execute & 56 & 36 \\ \hline
Other & Cases not fitting the above & 23 & 15 \\ \hline
\end{tabular}
\vspace{2mm}
\caption{Coarse-grained error distribution over 156 LLM-classified complete failures (primary label per case).}
\label{tab:error_groups} 
\end{table}

\newcommand{\iri}[1]{{\fontfamily{pcr}\selectfont\small #1}}
\vspace{2mm}
\begin{table}[!htbp]
\centering
\setlength{\tabcolsep}{2pt}%
\renewcommand{\arraystretch}{1.3}%
\scriptsize
\begin{tabular}{@{}>{\raggedright\arraybackslash}p{2cm}r>{\raggedright\arraybackslash}p{4.6cm}@{}}
\hline
\textbf{Category} & \textbf{$n$} & \textbf{Summarized example} \\ \hline
Wrong Class & 11 & used c:City instead of c:No\_Country\_Region \\ \hline
Wrong Property & 9 & used dblp:publishedInStream instead of dblp:publishedIn \\ \hline
Wrong Entity & 9 & used pid\_94:5174 instead of pid\_04:1616 (a different author) \\ \hline
\end{tabular}
\vspace{2mm}
\caption{Concept resolution errors (29 cases).}
\label{tab:error_concept}
\end{table}

\begin{table}[!htbp]
\centering
\setlength{\tabcolsep}{2pt}%
\renewcommand{\arraystretch}{1.3}%
\scriptsize
\begin{tabular}{@{}>{\raggedright\arraybackslash}p{2cm}r>{\raggedright\arraybackslash}p{4.6cm}@{}}
\hline
\textbf{Category} & \textbf{$n$} & \textbf{Summarized example} \\ \hline
Missing Constraint & 10 & omitted required COUNT(DISTINCT ?oceanTopic) \\ \hline
Incorrect Join & 10 & used service $\to$ schema:audienceType instead of service $\to$ schema:category $\to$ audience $\to$ schema:audienceType \\ \hline
\end{tabular}
\vspace{2mm}
\caption{Query construction errors (20 cases).}
\label{tab:error_construction}
\end{table}

\begin{table}[!htbp]
\centering
\setlength{\tabcolsep}{2pt}%
\renewcommand{\arraystretch}{1.3}%
\scriptsize
\begin{tabular}{@{}>{\raggedright\arraybackslash}p{2cm}r>{\raggedright\arraybackslash}p{4.6cm}@{}}
\hline
\textbf{Category} & \textbf{$n$} & \textbf{Summarized example} \\ \hline
Result Shape & 20 & used ?paper ?paperName ?metricCount instead of ?paperId ?metricCount \\ \hline
ASK vs.\ SELECT & 4 & used SELECT ?paper ?pid instead of ASK \\ \hline
IRI vs.\ Label & 4 & used "Ammonium Nitrate 34-0-0" instead of individuals:AmendmentType.30 \\ \hline
\end{tabular}
\vspace{2mm}
\caption{Answer representation errors (28 cases).}
\label{tab:error_repr}\vspace{2mm}
\end{table}

\begin{table}[!htbp]
\centering
\setlength{\tabcolsep}{2pt}%
\renewcommand{\arraystretch}{1.3}%
\scriptsize
\begin{tabular}{@{}>{\raggedright\arraybackslash}p{2cm}r>{\raggedright\arraybackslash}p{4.6cm}@{}}
\hline
\textbf{Category} & \textbf{$n$} & \textbf{Summarized example} \\ \hline
Agent Gave Up & 50 & Agent gave up after being mostly stuck in the search phase for 20 iterations \\ \hline
No Valid Query & 6 & The submitted query used \texttt{rdfs:label} without declaring the \texttt{rdfs:} prefix \\ \hline
\end{tabular}
\vspace{2mm}
\caption{Query generation/execution errors (56 cases).}
\label{tab:error_execution}\vspace{2mm}
\end{table}

\begin{table}[!htbp]
\centering
\setlength{\tabcolsep}{2pt}%
\renewcommand{\arraystretch}{1.3}%
\scriptsize
\begin{tabular}{@{}>{\raggedright\arraybackslash}p{2cm}r>{\raggedright\arraybackslash}p{4.6cm}@{}}
\hline
\textbf{Category} & \textbf{$n$} & \textbf{Summarized example} \\ \hline
Ambiguous Question & 10 & not clear how to rank the top 5 papers: all 4,482 papers tie at one distinct method \\ \hline
Top-$k$ / LIMIT Tie-Break & 3 & used ORDER BY DESC(?methodCount) ?name instead of ORDER BY DESC(?methodCount) \\ \hline
\end{tabular}
\vspace{2mm}
\caption{Other errors (23 cases). The remaining 10 cases fall into rare miscellaneous categories (one or two cases each) that are not shown.}
\label{tab:error_other}\vspace{2mm}
\end{table}

\par\noindent\begin{minipage}{\columnwidth}
\section{KGVoyager System Prompt and Tool Specification}
\label{sec:appendix}
\label{sec:appendix_prompt_kgvoyager_tools}

\begin{figure}[H]
\begin{tcolorbox}[colback=green!5!white, colframe=green!75!black,
                  unbreakable,nobeforeafter]
\footnotesize
You are a SPARQL query generation assistant. Your job is to generate an accurate SPARQL query that answers the user's question by dynamically discovering the knowledge graph's structure.

You operate in a think-act-observe loop: reason about what you need, invoke a tool, then interpret the results before deciding the next action. Follow this workflow:

\begin{enumerate}[leftmargin=*,topsep=2pt,itemsep=2pt]
\item \textbf{SEARCH:} Identify the classes, properties, and entities implied by the question. Search for candidate class and property URIs using natural-language terms. If the question mentions named entities, resolve them by first finding their class, then looking up instances by name. Use relevance scores and usage counts to select the most likely candidates.
\item \textbf{EXPLORE:} If the search results do not clearly reveal how properties connect entity types, explore property usage patterns to see actual domain/range types and example triples. If you need to understand the neighborhood of a specific entity, explore its relations.
\item \textbf{CONSTRUCT AND TEST:} Build the SPARQL query incrementally using the discovered URIs. Execute intermediate drafts to verify correctness. If results are unexpected, revisit earlier steps -- you may have the wrong property, class, or triple pattern. Once the query returns the expected results, call the answer function. If you cannot find a satisfactory query, call cancel.
\end{enumerate}

Keep in mind:
\begin{itemize}[leftmargin=*,topsep=2pt,itemsep=2pt]
\item Always strategically decide which tool call to make next. Include reasoning about your objective with the tool call whenever possible
\item Never make a redundant tool call. You will be highly penalized if you repeat a tool call or perform tool calls of very little value (eg, verifying something already obvious). It's absolutely normal if you only use a subset of the available tools to solve the problem.
\item Overall solving this with the least amount of tool calls but the perfect accuracy is the ultimate goal.
\item Never return a SPARQL query before verifying it is correct and returns results that answer the question.
\item For questions with a ``True'' or ``False'' answer the SPARQL query should be an ASK query.
\end{itemize}

Available knowledge graphs:\\
\texttt{\{\{ knowledge\_graphs \}\}}
\end{tcolorbox}
\caption{\kgv system prompt.}
\label{fig:kgv-prompt}
\end{figure}
\end{minipage}\par

\begin{figure}[H]
    \includegraphics[width=1.03\columnwidth]{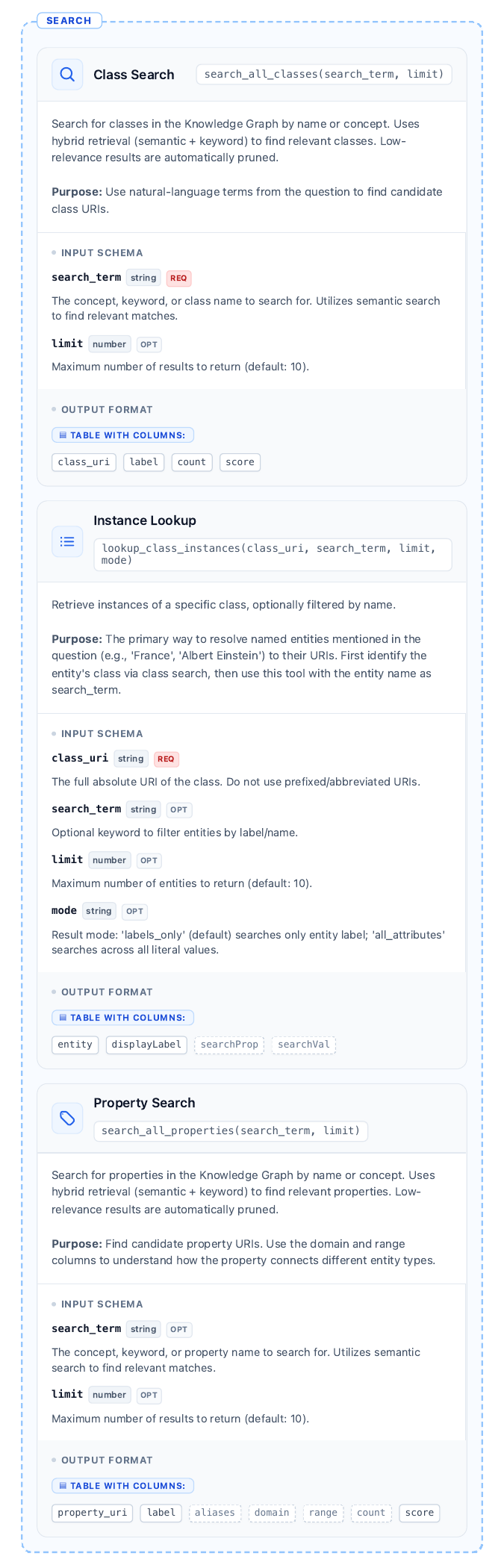}
    \caption{\kgv tool specification: search tools.}
\end{figure}
\begin{figure}[H]
    \includegraphics[width=1.06\columnwidth, trim=0cm 4.5cm 0cm 0cm, clip]{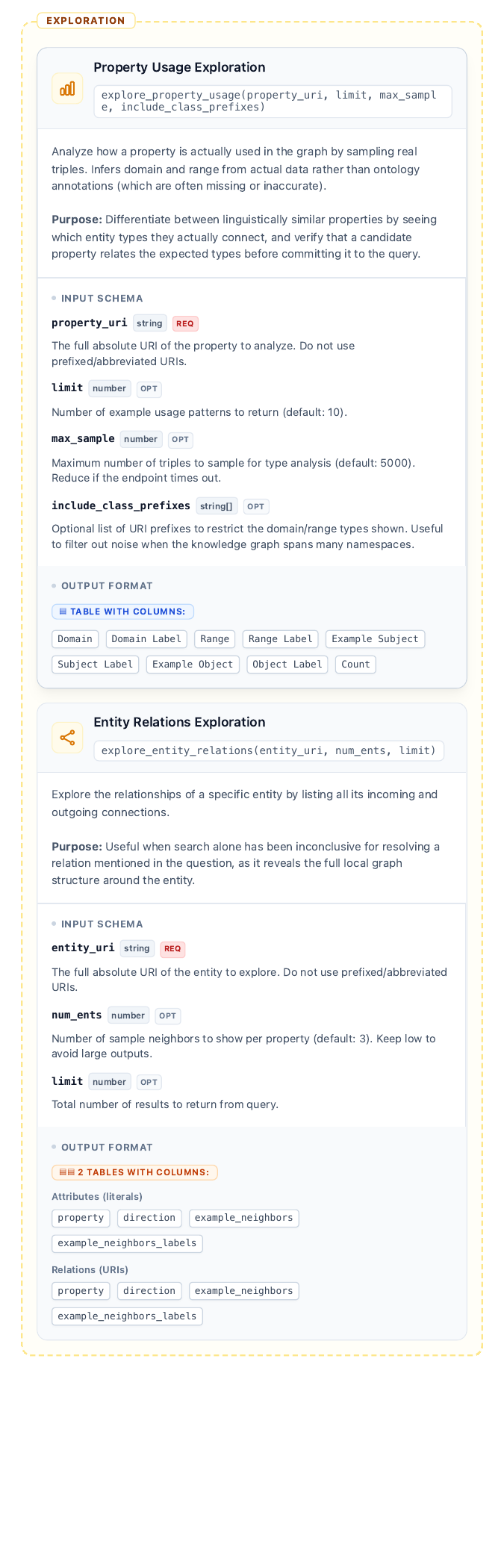}
    \caption{\kgv tool specification: exploration tools.}
\end{figure}
\begin{figure}[H]
    \includegraphics[width=1.06\columnwidth, trim=0cm 13cm 0cm 0cm, clip]{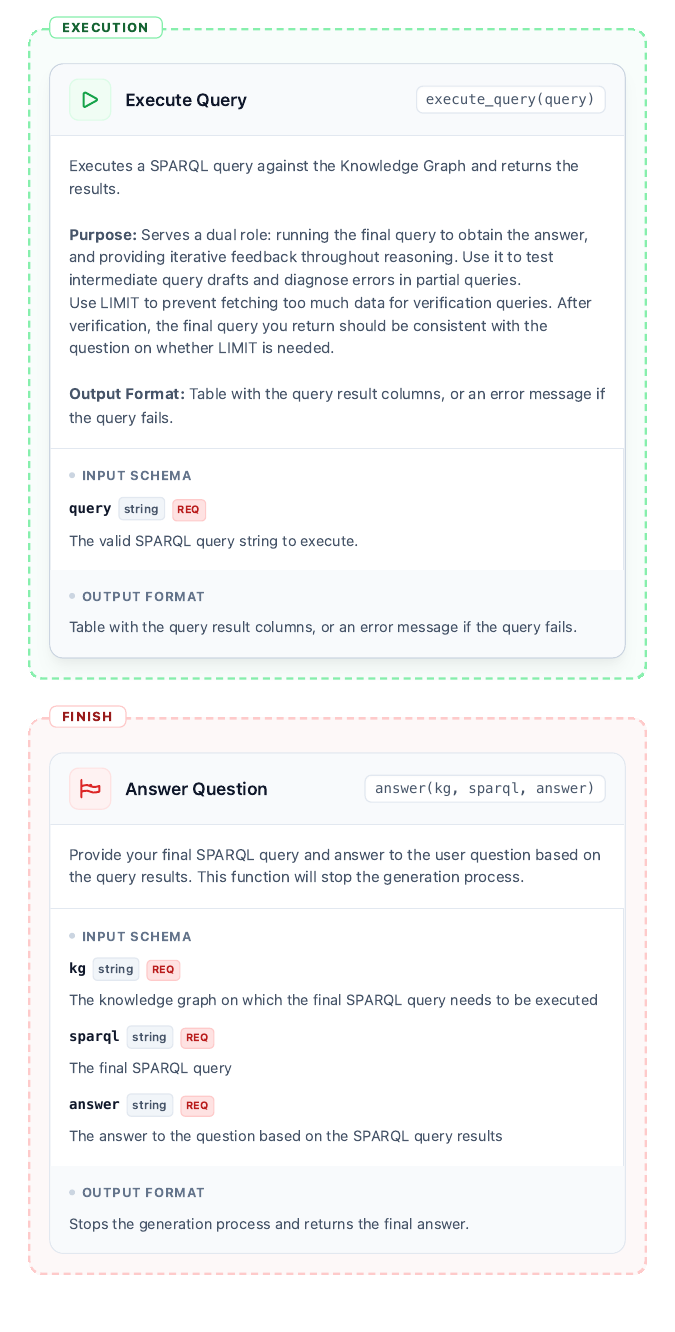}
    \caption{\kgv tool specification: execution tools.}
\end{figure}

\end{document}